\documentclass[12pt]{article}
\usepackage{amsmath}
\usepackage{graphicx}
\usepackage{geometry}
\usepackage[applemac]{inputenc}

\usepackage{natbib}

\ifx\pdfoutput\@undefined\usepackage[usenames,dvips]{color}
\else\usepackage[usenames,dvipsnames]{color}
\IfFileExists{pdfcolmk.sty}{\usepackage{pdfcolmk}}{} 
\fi
\usepackage[plainpages=false,pdfpagelabels,pagebackref=false,naturalnames=true,hyperindex=true,pdftitle={Information and Computation},pdfauthor={Carlos Gershenson}]{hyperref}
\hypersetup{colorlinks=true,
urlcolor=Cerulean,linkcolor=BrickRed,citecolor=RoyalBlue,
  pdfpagemode=None,
  pdfstartview=FitH}
\usepackage[all]{hypcap}

\begin{document}

\title{Information and Computation}
\author{Carlos Gershenson$^{1,2}$ \\
$^{1}$ Departamento de Ciencias de la Computaci\'on\\
Instituto de Investigaciones en Matem\'aticas Aplicadas y en Sistemas \\
Universidad Nacional Aut\'onoma de M\'exico\\
A.P. 20-726, 01000 M\'exico D.F. M\'exico\\
\href{mailto:cgg@unam.mx}{cgg@unam.mx} \
\url{http://turing.iimas.unam.mx/~cgg} \\
$^{2}$ Centro de Ciencias de la Complejidad \\
Universidad Nacional Aut\'onoma de M\'exico}
\maketitle

\begin{abstract}
In this chapter, concepts related to information and computation are reviewed in the context of human computation. A brief introduction to information theory and different types of computation is given. Two examples of human computation systems, online social networks and Wikipedia, are used to illustrate how these can be described and compared in terms of information and computation. 
\end{abstract}






\section{Introduction}

Before delving into the role of information theory as a descriptive tool for human computation~\citep{vonAhn2009HumanComputation}, we have to agree on at least two things: what is human, and what is computation, as human computation is at its most general level computation performed by humans. It might be difficult to define what makes us human, but for practical purposes we can take an ``I-know-it-when-I-see-it" stance. For computation, on the other hand, there are formal definitions, tools and methods that have been useful in the development of digital computers and can also be useful in the study of human computation.


\section{Information}

Information has had a long and interesting history~\citep{gleick2011information}. It was Claude Shannon~\citeyearpar{Shannon1948} who developed mathematically the basis of what we now know as \emph{information theory}~\citep{Ash1990Information}. Shannon was interested in particular on how a message could be transmitted reliably across a noisy channel. This is very relevant for telecommunications. Still, information theory has proven to be useful beyond engineering~\citep{vonBaeyer2004}, as  anything can be described in terms of information~\citep{Gershenson:2007}. 

A brief technical introduction to Shannon information $H$ is given in Appendix \ref{app:info}. The main idea behind this measure is that messages will carry more information if they reduce uncertainty. Thus, if some data is very regular, i.e. already certain, more data will bring new information, so $H$ will be low, i.e. few or no new information. If data is irregular or close to random, then more data will be informative and $H$ will be high, since this new data could not have been expected from previous data.

Shannon information assumes that the meaning or decoding is fixed, and this is generally so for information theory. The study of meaning has been made by semiotics~\citep{Peirce:1991,Eco:1979}. The study of the evolution of language~\citep{Christiansen2003Language-evolut} has also dealt with how meaning is acquired by natural or artificial systems~\citep{steels1997synthetic}.

Information theory can be useful for different aspects of human computation. It can be used to measure, among other properties: the information transmitted between people, novelty, dependence, and complexity~\citep{ProkopenkoEtAl2007,GershensonFernandez:2012}. For a deeper treatment of information theory, the reader is referred to the textbook by Cover and Thomas~\citeyearpar{CoverThomas2006}.

\section{Computation}

Having a most general view, computation can be seen simply as the transformation of information~\citep{Gershenson:2007}. If anything can be described in terms of information, then anything humans do could be said to be human computation. However, this notion is too broad to be useful.

A formal definition of computation was proposed by Alan Turing~\citeyearpar{Turing:1936}. He defined an abstract ``machine" (a Turing machine) and defined ``computable functions" as those which the machine could calculate in finite time. This notion is perhaps too narrow to be useful, as Turing machines are cumbersome to program and it is actually debated whether Turing machines can model all human behavior~\citep{Edmonds:2012}.

An intermediate and more practical notion of computation is \emph{the transformation of information by means of an algorithm or program.} This notion on the one hand tractable, and on the other hand is not limited to abstract machines.

In this view of computation, the algorithm or program (which can be run by a machine or animal) defines rules by which information will change. By studying at a general level what happens when the information introduced to a program (input) is changed, or how the computation (output) changes when the program is modified (for the same input), different types of dynamics of information can be identified:

\begin{description}
\item[Static. ] Information is not transformed. For example, a crystal has a pattern which does not change in observable time.
\item[Periodic. ] Information is transformed following a regular pattern. For example, planets have regular cycles which in which information measured is repeated every period.
\item[Chaotic. ] Information is very sensitive to changes to itself or the program, it is difficult to find patterns. For example, small changes in temperature or pressure can lead to very different meteorological futures, a fact which limits the precision of weather prediction.
\item[Complex. ] Also called critical, it is regular enough to preserve information but allows enough flexibility to make changes. It balances robustness and adaptability~\citep{Langton1990}. Living systems would fall in this category.
\end{description}
	
Wolfram~\citeyearpar{Wolfram2002}  conjectured that there are only two types of computation: universal or regular. In other words, programs are either able to perform any possible computation (universal), or they are simple and limited (regular). This is still an open question and the theory of computation is an active research area.

\section{Computing Networks}

Computing networks (CNs) are a formalism proposed to compare different types of computing structures~\citep{Gershenson:2010b}. CNs will be used to compare neural computation (information transformed by neurons), machine distributed computation (information transformed by networked computers), and human computation.

In computing networks, nodes can process information (compute) and exchange information through their edges, each of which connects the output of node with the input of another node. 
A computing network 
is defined as \textbf{a set of nodes $N$ linked by a set of edges $K$ used by an algorithm $a$ to compute a function $f$}~\citep{Gershenson:2010b}. Nodes and edges can have internal variables that determine their state, and functions that determine how their state changes. CNs can be stochastic or deterministic, synchronous or asynchronous, discrete or continuous.

\begin{figure}[htbp]
\begin{center}
\includegraphics[width=0.5\textwidth]{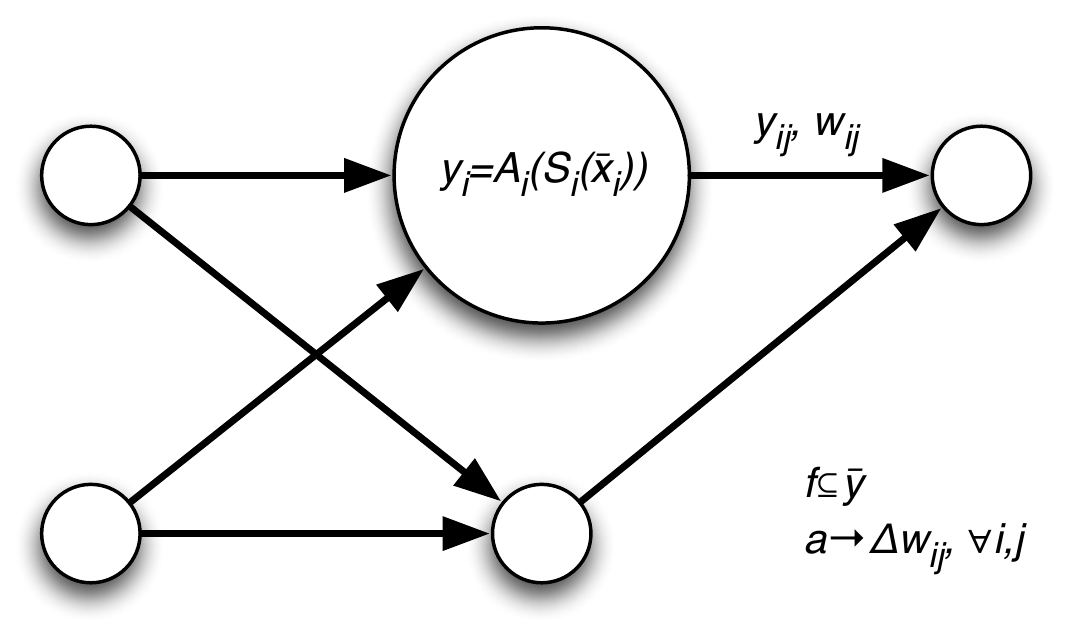}
\caption{A NN represented as a CN. }
\label{fig:NN}
\end{center}
\end{figure}

In a CN description of a \textbf{neural network} (NN) model, \emph{nodes} represent neurons. Each neuron $i$ has a continuous state (output) determined by a function $y_i$ which is composed by two other functions: the weighted sum $S_i$ of its inputs $\bar{x}_i$ and an activation function $A_i$, usually a sigmoid. Directed \emph{edges} $ij$ represent synapses, relating outputs $y_i$ of neurons $i$ to inputs $x_j$ of neurons $j$, as well as external inputs and outputs with the network. Edges have a continuous state $w_{ij}$ (weight) that relates the states of neurons. The \emph{function} $f$ may be given by the states of a subset of $N$ (outputs $\bar{y}$), or by the complete set $N$. NNs usually have two dynamical scales: a ``fast" scale where the network function $f$ is calculated by the functional composition of the function $y_i$ of each neuron $i$, and a ``slow" scale where a learning \emph{algorithm} $a$ adjusts the weights $w_{ij}$ (states) of edges. 
There is a broad diversity of algorithms $a$ used to update weights in different types of NN. Figure \ref{fig:NN} illustrates NNs as CNs.

\begin{figure}[htbp]
\begin{center}
\includegraphics[width=0.5\textwidth]{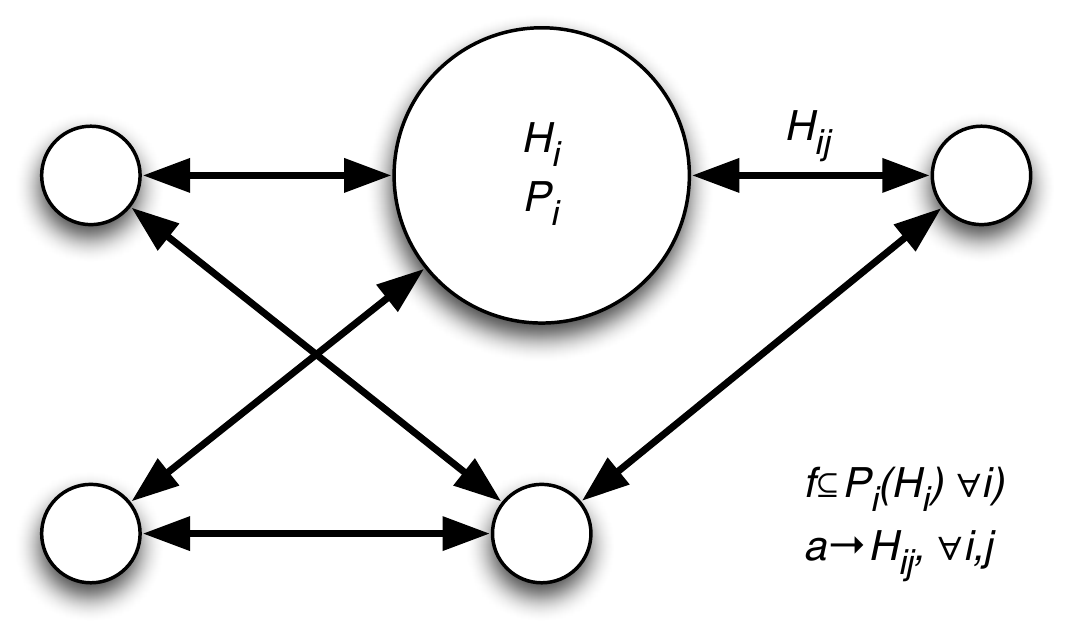}
\caption{A DC system or a HC system represented as a CN. }
\label{fig:DC}
\end{center}
\end{figure}

Digital machines carrying out \textbf{distributed computation} (DC) can also be represented as CNs.
\emph{Nodes} represent computers while \emph{edges} represent network connections between them. Each computer $i$ has information $H_i$ which is modified by a program $P_i(H_i)$. Physically, both $H_i$ and $P_i$ are stored in the computer memory, while the information transformation is carried out by a processor. Computers can share information $H_{ij}$ across edges using a communication protocol. The \emph{function} $f$ of the DC will be determined by the output of $P_i(H_i)$ of some or all of the nodes, which can be seen as a ``fast" scale. Usually there is an algorithm $a$ working at a ``slower" scale, determining and modifying the interactions between computers, i.e. the network topology. Figure \ref{fig:DC} shows a diagram of DC as a CN.

\textbf{Human computation} (HC) can be described as a CN in a very similar way than DC. People are represented as \emph{nodes} and their interactions as \emph{edges}. People within a HC system transform information $H_i$ following a program $P_i(H_i)$. In many cases, the information shared between people $H_{ij}$ is transmitted using digital computers, e.g. in social networks, wikis, forums, etc. In other cases, e.g. crowd dynamics, information $H_{ij}$ is shared through the environment: acustically, visually~\citep{Moussaid:2011}, stigmergically~\citep{Doyle:2013}, etc. The function $f$ of a HC system can be difficult to define, since in many cases the outcome is observed and described only \emph{a posteriori}. Still, we can say that $f$ is a combination of the computation carried out by people. An algorithm $a$ would determine how the social links change in time. Depending on the system, $a$ can be slower than $f$ or vice versa.

In DC, the algorithm $a$ is centrally determined by a designer, while in most HC systems, the $a$ is determined and executed by people (nodes) themselves.

Using information theory, we can measure how much information $H_{ij}$ is transmitted between people, how much each person receives and produces, and how much the entire system receives and produces. In many cases, machines enable this transmission and thus also facilitate its measurement. Comparing the history of information transfers and current information flows can be used to measure the novelty in current information.

\section{Examples}

\subsection{Social Networks}

A straightforward example of human computation can be given with online social networks. There are key differences, e.g. links are bidirectional in Facebook (my friends also have me as their friend) and unidirectional in Twitter (the people I follow do not necessarily follow me, I do not necessarily follow my followers). People and organizations are represented with their accounts in the system as nodes, and they receive information through their incoming links, They can share this information with their outgoing links and also produce novel information that their links may receive. People can decide how to create or eliminate social links, i.e. $a$ is decided by individuals. 

These simple rules of the information dynamics on social networks are able to produce very interesting features of human computation~\citep{Lerman2010Information-con}, which can be described as functions $f$. For example, non-official news can spread very quickly through social networks, challenging mass media dominated by some governments. On the other hand, false rumors can also spread very quickly, potentially leading to collective misbelief. Nevertheless, it has been found that the dynamics of false rumors spreading is different from that of verifiable information~\citep{Castillo2011Information-cre}.

Describing social networks as CNs is useful because interactions are stated explicitly. Moreover, one can relate different scales with the same model: local scale (nodes), global scale (networks), and meso scales (modules); and also temporal scales: fast ($f$) and slow ($a$). Information theory can be used to detect novelty in social interactions (high $H$ values in edges), imitation (low $H$ values in edges), unusual patterns (``fake" information), correlations (with mutual information), and communities (modules~\citep{Newman:2010}).

\subsection{Wikipedia}

Wikipedia gives a clear example of the power of human computation. Millions of people (nodes) from all over the world have collaboratively built the most extensive encyclopedia ever. The sharing of information is made through editable webpages on a specific topic. Since these pages can potentially link more than two people (editing the webpage), the links can be represented as those of a hypernetwork~\citep{Johnson2009Hypernetworks-i}, where edges can link more than two nodes (as in usual networks). The information in pages (hyperedges) can be measured, as it changes over time with the editing made by people linked to them. The information content delivered by different authors can be measured with $H$. When this is increased, it implies novelty. The complexity of the webpages, edits, and user interactions can also be measured, seen as a balance between maximum information (noise) and minimum information (stasis)~\citep{Fernandez2013Information-Mea}. 

The function $f$ of Wikipedia is its own creation, growth, and refinement: the pages themselves are the output of the system. Again, people decide which pages to edit, so the algorithm $a$ is also decided by individuals.

Traditionally, Wikipedia---like any set of webpages---is described as a network of pages with directional edges from pages that link to other pages. This is a useful description to study the structure of Wikipedia itself, but it might not be the most appropriate in the context of human computation, as no humans are represented. Describing Wikipedia as a CN, the relationships between humans and the information they produce collaboratively is explicit, providing a better understanding of this collective phenomenon.



\section{Conclusions}

Concepts related to information and computation can be applied to any system, as anything can be described in terms of information~\citep{Gershenson:2007}. Thus, HC can also benefit from the formalisms and descriptions related to information and computation.
 
CNs are general, so they can be used to describe and compare any HC system.
For example, it is straightforward to represent online social networks such as Facebook, Twitter,  LinkedIn, Google+, Instagram, etc. as CNs. As such, their structure, functions, and algorithms can be contrasted, and their local and global information dynamics can be measured. The properties of each of these online social networks could be compared with other HC systems, such as Wikipedia. 

Moreover, CNs and Information Theory can be used to design and self-monitor HC systems~\citep{GershensonDCSOS}. For example, information overload should be avoided in HC systems. The formalisms presented in this chapter and in the cited material can be used to measure information inputs, transfers, and outputs to avoid not only information overload, but also information poverty~\citep{Bateson1972}. 

In our age where data is overflowing, we require appropriate measures and tools to be able to make sense out of ``big data". Information and computation provide some of these measures and tools. There are still several challenges and opportunities ahead, but what has been achieved so far is very promising and invites us to continue exploring appropriate descriptions of HC systems.

\section*{Acknowledgments}

I should like to thank Matthew Blumberg and Pietro Michelucci for useful advice. 
This work was partially supported by SNI membership 47907 of CONACyT, Mexico. 

\appendix

\section{Shannon Information}
\label{app:info}

\begin{figure}[htbp]
\begin{center}
  \includegraphics[width=0.5\textwidth]{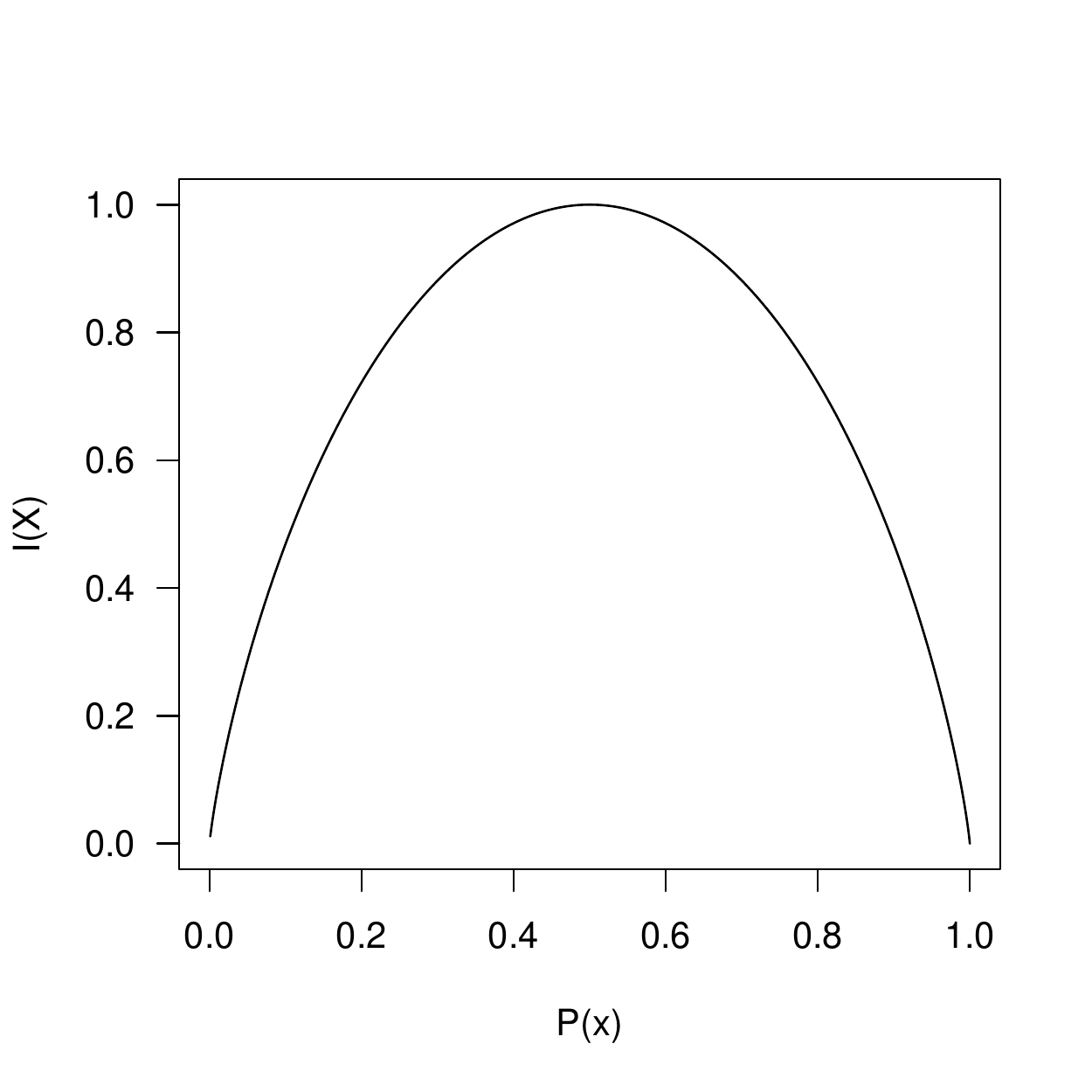}\\
\caption{Shannon's Information $H(X)$ of a binary string $X$ for different probabilities $P(x)$. Note that $P(0)=1-P(1)$.}
\label{fig:entropy}
\end{center}
\end{figure}

Given a string $X$, composed by  a sequence of values $x$ which follow a probability distribution $P(x)$, information (according to Shannon) is defined as:

\begin{equation}
H=-\sum{P(x) \log P(x)}.
\label{eq:I}
\end{equation}
For binary strings, the most commonly used in ICT systems, the logarithm is usually taken with base two. For example, if the probability of receiving ones is maximal ($P(1)=1$) and the probability of receiving zeros is minimal ($P(0)=0$), the information is minimal, i.e. $H=0$, since we know beforehand that the future value of $x$ will be $1$. Information is zero because future values of $x$ do not add anything new, i.e. the values are known beforehand. If we have no knowledge about the future value of $x$, as with a fair coin toss, then $P(0)=P(1)=0.5$. In this case, information will be maximal, i.e. $H=1$, because a future observation will give us all the relevant information, which is also independent of previous values. Equation \ref{eq:I} is plotted in Figure \ref{fig:entropy}. Shannon information can be seen also as a measure of uncertainty. If there is absolute certainty about the future of $x$, be it zero ($P(0)=1$) or one ($P(1)=1$), then the information received will be zero. If there is no certainty due to the probability distribution ($P(0)=P(1)=0.5$), then the information received will be maximal. Shannon used the letter $H$ because equation \ref{eq:I} is equivalent to Boltzmann's entropy in thermodynamics, which is also defined as $H$.
The unit of information is the bit. One bit represents the information gained when a binary random variable becomes known. 

A more detailed explanation of information theory, as well as measures of complexity, emergence, self-organization, homeostasis, and autopoiesis based on information theory can be found in~\citet{Fernandez2013Information-Mea}.

\bibliographystyle{cgg}

\bibliography{carlos,sos,RBN,complex,information,COG,traffic}

\end{document}